\documentclass[sigconf]{acmart}
\usepackage[linesnumbered,ruled,vlined]{algorithm2e}
\usepackage{multirow}
\usepackage{url}
\usepackage{afterpage}
\usepackage{xcolor}
\usepackage{placeins}  
\usepackage{multicol}
\usepackage{booktabs}
\usepackage[utf8]{inputenc}

\usepackage{tcolorbox} 
\usepackage{balance}

\newcommand{\huang}[1]{{\color{black}{#1}}}
\newcommand{\zhou}[1]{{\color{black}{#1}}}

\AtBeginDocument{%
  }

\copyrightyear{2026}
\acmYear{2026}
\setcopyright{cc}
\setcctype{by}
\acmConference[KDD '26]{Proceedings of the 32nd ACM SIGKDD Conference on Knowledge Discovery and Data Mining V.2}{August 09--13, 2026}{Jeju Island, Republic of Korea}
\acmBooktitle{Proceedings of the 32nd ACM SIGKDD Conference on Knowledge Discovery and Data Mining V.2 (KDD '26), August 09--13, 2026, Jeju Island, Republic of Korea}
\acmDOI{10.1145/3770855.3817586}
\acmISBN{979-8-4007-2259-2/2026/08}

\begin{document}


\title[Clarify-Then-Search:
A Clarification Benchmark for Deep Search with End-to-End Nugget Restoration]{Clarify-Then-Search: A Clarification Benchmark for Deep Search with End-to-End Nugget Restoration}
\author{Deqiang Huang}
\authornote{This work was done when the first author was an intern in the Big Model Frontier Research Department of Baidu Inc., under the supervision of Jingbo Zhou.}
\affiliation{%
  \institution{University of Science and Technology of China}
  \city{Hefei}
  \state{Anhui}
  \country{China}}
\email{deqianghuang@mail.ustc.edu.cn}

\author{Jingbo Zhou}
\authornote{Corresponding authors.}
\affiliation{%
  \department{Frontier Research Department}
  \institution{Baidu Inc.}  
  \city{Beijing}
  \country{China}}
\email{zhoujingbo@baidu.com}

\author{Xinjiang Lu}
\affiliation{%
  \department{Frontier Research Department}
  \institution{Baidu Inc.}
  \city{Beijing}
  \country{China}}
\email{luxinjiang@baidu.com}

\author{Tong Xu}
\authornotemark[2] 
\affiliation{%
  \institution{University of Science and Technology of China}
  \city{Hefei}
  \state{Anhui}
  \country{China}}
\email{tongxu@ustc.edu.cn}

\author{Hua Wu}
\affiliation{%
  \department{Frontier Research Department}
  \institution{Baidu Inc.}
  \city{Beijing}
  \country{China}}
\email{wu_hua@baidu.com}

\author{Enhong Chen}
\authornotemark[2] 
\affiliation{%
  \institution{University of Science and Technology of China}
  \city{Hefei}
  \state{Anhui}
  \country{China}}
\email{cheneh@ustc.edu.cn}
\renewcommand{\shortauthors}{Deqiang Huang et al.}

\begin{abstract}
Deep search is brittle on underspecified user queries: missing constraints (e.g., time, location, scope, or definitions) often lead to retrieval drift and incomplete answers. Although LLMs can ask clarifying questions, existing evaluations of clarification for deep search remain limited.
We introduce a \emph{clarify-then-search} benchmark \zhou{to evaluate the ability of LLMs to ask clarifying questions that facilitate} deep search. 
\zhou{Built on real-world query data from the Baidu search engine}, the benchmark contains 518 curated instances that target the \zhou{scenarios} where clarification is most needed: each instance provides an intent query $q^{*}$ (\texttt{fused\_query}) and a corresponding underspecified query $q$ (\texttt{blurred\_query}). We run a deep-search backbone (we use WebDancer in this study) \emph{once} per $q^{*}$ to archive evidence and construct a \emph{static} golden reference as weighted, evidence-grounded nuggets with traceable source identifiers. At evaluation time, a Clarifier asks $k\!\in\!\{1,2,3\}$ questions; a closed-book User Answerer replies using only information explicitly stated in $q^{*}$ (otherwise returning \texttt{unknown}); and a closed-book Rewriter produces a rewritten query $\hat{q}$ using only $q$ and the elicited Q\&A pairs. WebDancer then executes on $\hat{q}$, and we score end-to-end utility by \texttt{restore\_score\_100} (0--100),\huang{ a weighted nugget-recall score with partial credit against the static gold.} 

\huang{Across all evaluated models, clarification improves over the no-interaction baseline at $k=1$, and larger budgets ($k=2,3$) yield further gains. Notably, GPT-5.2 achieves the highest mean score at $k=1$, while ERNIE-4.5-Turbo-128K leads the evaluated open-weight models at $k=1$ and becomes the overall top-performing model at $k=3$, surpassing GPT-5.2, Claude-Sonnet-4.5, and Gemini-2.5-Pro. Interaction diagnostics further reveal a consistent failure mode under strict closed-book clarification: many systems over-ask region-only questions that are often unanswerable from the intent and thus elicit \texttt{unknown}, producing low-yield interactions despite additional turns. Overall, the benchmark enables leakage-resistant and reproducible evaluation of clarify-then-search pipelines, while supporting fine-grained analyses of question utility, answerability, and clarification budget effects in deep search.} All data and code are available on the project page: \url{https://clarify-then-search.github.io/}.
\end{abstract}



\maketitle
\section{Introduction}
\label{sec:intro}

Deep search systems are increasingly deployed for complex information needs, yet they remain brittle when user queries are ambiguous or underspecified.
In real search logs, users frequently omit critical constraints such as temporal scope, geographic region, entity disambiguation, numeric thresholds, or even the intended definition of a concept.
Compared with standard retrieval, underspecification in deep search has larger downstream consequences: an early ambiguity can mislead search planning, waste tool-call budget on low-yield exploration, trigger off-target evidence visits, and propagate into incomplete or incorrect final answers.
This motivates a \emph{clarify-then-search} paradigm: ask targeted clarification questions, incorporate user feedback into a rewritten query, and then execute deep search more efficiently and accurately.

Despite growing interest in clarification question generation, progress is bottlenecked by evaluation.
First, many prior benchmarks focus on question-level properties, such as fluency, plausibility, or topical relevance, which are weak proxies for whether clarification actually improves downstream search utility.
Second, standard retrieval-oriented evaluation captures only part of the effect of clarification.
In deep search, clarification affects not only document recall, but also planning, evidence browsing, aggregation, and final answer completeness over multi-faceted information needs.
Third, end-to-end evaluations can suffer from \emph{intent leakage}: rewriting or retrieval stages may indirectly observe oracle intent, such as the original clear query or ground-truth slots, making it difficult to attribute gains to clarification quality.

As a result, it remains unclear whether stronger clarification behavior translates into measurable improvements under realistic information constraints.
To address these issues, our benchmark (i) evaluates the clarification ability of LLMs by downstream utility in an end-to-end clarify$\rightarrow$rewrite$\rightarrow$deep-search pipeline, (ii) enforces a closed-book protocol to prevent intent leakage, and (iii) uses a static nugget-based gold target to enable reproducible comparisons.
In this benchmark, \emph{information gain} is used in an operational sense: a clarification question is useful if it elicits intent-grounded information that can be safely incorporated into the rewrite and ultimately improves downstream nugget restoration.

Based on real-world queries from the Baidu search engine, we introduce a fully open benchmark\footnote{Project page: \url{https://clarify-then-search.github.io/}.} that evaluates clarification by \emph{end-to-end utility} in a fixed deep-search pipeline.
In this study, the pipeline is powered by WebDancer~\cite{wu2025webdancer}, although other deep-search backbones can also be used.
Our benchmark contains 518 curated instances that target scenarios where clarification is most needed: each instance includes an underspecified query (\texttt{blurred\_query}) paired with its underlying intent query (\texttt{fused\_query}).
For each intent, we build a \emph{static} golden reference by running WebDancer once on \texttt{fused\_query} and extracting evidence-grounded nuggets traceable to retrieved sources.
At evaluation time, a Clarifier generates $k$ clarification questions ($k\in\{1,2,3\}$), receives constrained answers from a closed-book user simulator, rewrites the query, and runs WebDancer on the rewritten query.
We measure end-to-end utility by how well the final output restores the static nuggets, reported as \texttt{restore\_score\_100}, i.e., weighted nugget recall (0--100) with partial credit for \texttt{partial} coverage (Section~\ref{sec:eval}).

A central design goal is to prevent intent leakage while preserving a realistic interaction pattern for LLM clarification.
We therefore adopt a two-phase closed-book protocol.
The \textbf{User Answerer (UA)} sees only \texttt{fused\_query} and the current clarification question, and must answer \emph{only what is asked}; if the intent does not explicitly specify the requested attribute, UA outputs \texttt{unknown}.
The \textbf{Rewriter} sees only \texttt{blurred\_query} and the clarification Q\&A pairs, and must rewrite without introducing new entities or constraints; when an answer is \texttt{unknown}, it must not be forcefully concretized.
This separation ensures that any additional constraint in the rewritten query must come from information actually elicited through clarification, rather than being copied from the hidden intent.
It also exposes intermediate interaction signals, such as answerability, multi-turn no-signal probability, and question-type bias, for diagnosis beyond the final score.

Our results show that (i) all Clarifiers improve over the no-interaction baseline at $k=1$, and (ii) larger clarification budgets generally yield further gains.
We further observe that GPT-5.2 achieves the highest mean \texttt{restore\_score\_100} with a single clarification question ($k=1$). Among the evaluated open-weight models, ERNIE-4.5-Turbo-128K leads at $k=1$, outperforming Qwen3-235B-A22B-Instruct, Kimi-K2-Instruct, and DeepSeek-V3.2. ERNIE-4.5-Turbo-128K further becomes the overall top-performing model at $k=3$, surpassing GPT-5.2, Claude-Sonnet-4.5, and Gemini-2.5-Pro.

However, intermediate diagnostics reveal systematic failure modes, especially a bias toward region questions that often elicit \texttt{unknown} under the closed-book constraint, showing that reliably improving deep search through clarification remains challenging.
Additional robustness analysis further shows that while absolute scores vary across coverage judges, the main improvement trend remains stable across ERNIE-, GPT-, and Claude-based judges.

Our contributions are summarized as follows:
\begin{itemize}
    \item \textbf{Dataset and Task.} We formalize deep-search clarification as an end-to-end benchmark from underspecified queries to nugget restoration, and release a dataset of 518 curated instances derived from real-world Baidu search queries with standardized evaluation outputs.
    \item \textbf{Leakage-Free Protocol.} We introduce a two-phase closed-book evaluation protocol that prevents rewrite-stage exposure to oracle intent, enabling controlled comparisons of Clarifiers for $k\in\{1,2,3\}$.
    \item \textbf{Static Golden Nuggets and Metric.} For each intent query, we archive an evidence-grounded static golden reference with weighted nuggets and score systems by weighted nugget recall with partial credit (\texttt{restore\_score\_100}).
    \item \textbf{Diagnostic Signals.} The benchmark makes intermediate interaction behavior measurable, including UA answerability, multi-turn all-\texttt{unknown} rate, and question-type stratification, enabling analyses of why clarification succeeds or fails beyond final scores.
    \item \textbf{Reproducibility Artifacts.} We release the dataset, pipeline scripts, prompts, and per-instance tool traces/snippets needed to reproduce golden construction and evaluation against a fixed target.
\end{itemize}

\section{Related Work}
\label{sec:related}

\textbf{Clarifying Questions in Conversational Search.}
Prior work studies how systems ask clarifying questions when an initial query is underspecified.
Benchmarks and datasets such as Qulac and MIMICS support modeling clarification behavior and its impact on retrieval \cite{Aliannejadi2019Clarify,Zamani2020Mimics,Zamani2020ClarifyIR}.
Recent resources further broaden this line with challenge-style clarification datasets and offline/online evaluation protocols for multi-turn clarification \cite{Aliannejadi2020ClariQ,Tavakoli2022MIMICSDuo}.
Other work also investigates when and how to ask (or avoid) clarifying questions \cite{huang-etal-2026-uncertainty-aware}, including the role of negative feedback and interaction signals \cite{Bi2021CQNegative}.
These efforts often emphasize question quality or short-horizon retrieval effects.
In contrast, our benchmark evaluates clarification by \emph{end-to-end utility} (clarify $\rightarrow$ rewrite $\rightarrow$ deep search $\rightarrow$ nugget restoration) under a controlled protocol.

\textbf{Query Rewriting for Conversational Search.}
Query rewriting converts contextual or ambiguous inputs into explicit stand-alone queries and is central to conversational search.
TREC CAsT operationalizes this setting and provides rewritten queries for evaluation \cite{Dalton2020CAsT19}.
Learning-based approaches optimize rewriting via supervision or retrieval-oriented objectives, and LLM-based prompting can produce competitive rewrites \cite{wu-etal-2022-conqrr,ye-etal-2023-enhancing}.
Widely used rewriting datasets (e.g., CANARD and QReCC) further support modeling context-to-query reformulation and retrieval-oriented rewriting behavior \cite{Elgohary2019CANARD,Anantha2021QReCC}.
Many setups allow the rewriter to access the full original context, which can introduce intent leakage.
Our two-phase closed-book design prevents this: the rewriter never observes the intent query, isolating the contribution of clarification signals.

\textbf{Tool-Augmented Agents and Deep Search Evaluation.}
Tool-augmented agents that interleave reasoning with actions (e.g., web search \cite{wu2025webdancer,huang-etal-2025-manusearch,luo2025imagescope} and web agents \cite{he2024webvoyager,zhang2026omegause}) motivate new evaluation paradigms.
ReAct formalizes reasoning--acting interleaving \cite{Yao2023ReAct}, while benchmarks such as BrowseComp and API-Bank evaluate browsing or tool-use capabilities \cite{Wei2025BrowseComp,Li2023APIBank}.
Complementary agent benchmarks evaluate realistic web interaction and general LLM-as-agent competence in interactive environments \cite{Zhou2023WebArena,Yao2022WebShop,Liu2023AgentBench}.
Other tool-use benchmarks systematize large tool collections and evaluate whether LLMs can select and invoke tools reliably \cite{Qin2023ToolBench}.
Our focus is narrower but practically important: \emph{clarify-then-search} under strict information constraints.
We evaluate how well a system elicits and uses clarifying information to improve a query for a fixed deep-search backend (WebDancer), enabling controlled comparisons across Clarifiers.

\textbf{Nugget-Based and Evidence-Grounded Evaluation.}
Nugget-based protocols score coverage of atomic information units and have been used in QA-style evaluations \cite{Lin2006Pourpre}, with pyramid weighting improving robustness \cite{Dang2007Pyramid}.
This family of methods is closely related to TREC-style evaluation practices that emphasize coverage of key facts rather than surface-form matching \cite{Voorhees2004TRECQA}.
Recent RAG evaluation work revisits nugget extraction and matching as a fine-grained measure of answer completeness \cite{Pradeep2024AutoNugget}.
Our benchmark follows this direction but adds two properties:
(i) \emph{static golden nuggets} archived per intent query from a one-time WebDancer run for reproducibility, and
(ii) a restoration metric that measures end-to-end clarification utility by quantifying how much of the evidence-grounded golden nugget set is recovered.

\textbf{Summary.}
Existing work provides foundations for clarification, rewriting, tool-augmented search, and nugget-based evaluation.
We integrate these into a leakage-resistant, end-to-end benchmark for deep search, evaluated on a curated set of 518 underspecified queries, where clarification is scored by downstream nugget restoration against static, evidence-grounded gold under a closed-book interaction protocol.
\section{Task Definition and Closed-Book Protocol}
\label{sec:task}

\begin{figure*}[t]
  \centering
  \includegraphics[width=\textwidth]{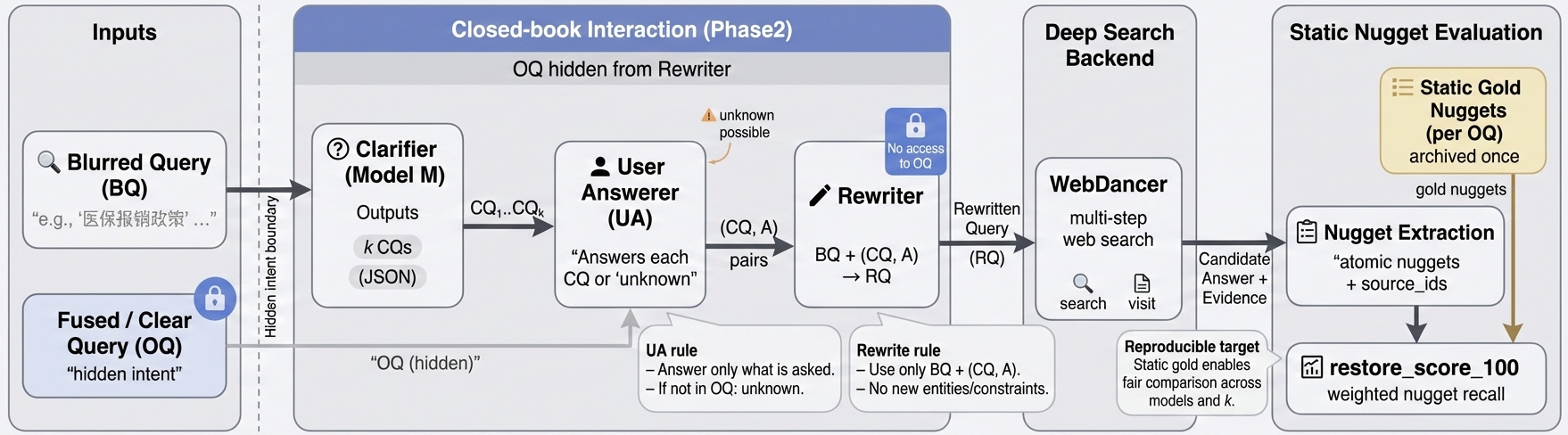}
  \caption{\textbf{Phase2 closed-book clarify-then-search pipeline.}
Given a blurred query (BQ), a Clarifier generates $k\in\{1,2,3\}$ clarification questions.
A constrained User Answerer (UA) replies using only the hidden intent query (OQ; \texttt{fused\_query}) and outputs \texttt{unknown} when the intent does not specify the requested attribute.
A Rewriter then produces a rewritten query (RQ; $\hat{q}$) from BQ and the Q\&A pairs without access to OQ.
WebDancer runs deep search on RQ, and performance is scored by \texttt{restore\_score\_100} against static golden nuggets archived once per OQ.}
  \label{fig:pipeline}
\end{figure*}

\subsection{Problem Setup}

Our benchmark consists of 518 instances, each defined by an intent query $q^{*}$ (stored as \texttt{fused\_query}) and a corresponding underspecified query $q$ (stored as \texttt{blurred\_query}).
Given only $q$, a \emph{Clarifier} produces $k$ clarification questions $C=(c_{1},\ldots,c_{k})$, where $k\in\{1,2,3\}$.
A constrained user then provides answers $A=(a_{1},\ldots,a_{k})$, and a \emph{Rewriter} generates a rewritten query $\hat{q}$ that should be more specific and retrieval-ready.
Finally, a fixed deep-search backend, WebDancer, runs on $\hat{q}$ to produce an answer with supporting evidence traces, and performance is evaluated against a static golden reference constructed for $q^{*}$.
Figure~\ref{fig:pipeline} summarizes the end-to-end evaluation pipeline.

\paragraph{Baselines and targets.}
We use two fixed reference points.
\textbf{Gold} corresponds to executing WebDancer once on the intent query $q^{*}$, archiving the resulting tool traces, and extracting evidence-grounded nuggets as a static reference.
\textbf{Orig} denotes executing WebDancer directly on the underspecified query $q$ without clarification and rewriting.
The benchmark therefore measures whether clarification and closed-book rewriting improve nugget restoration beyond \textbf{Orig}, using \textbf{Gold} as the fixed evaluation target.

\subsection{Two-Phase Closed-Book Interaction}

To prevent intent leakage in rewriting, we use two closed-book agents: a \emph{User Answerer} and a \emph{Rewriter}.
The key idea is to separate the component that can observe the hidden intent from the component that constructs the final query.
UA may access $q^{*}$ only to answer the current clarification question, while the Rewriter never observes $q^{*}$ and can only use information explicitly elicited through Q\&A.

\paragraph{User Answerer (UA)}
UA receives the intent query $q^{*}$ and a single clarification question $c_i$.
It must answer using only information \emph{explicitly} present in $q^{*}$ and must not introduce new entities, facts, or constraints.
If $q^{*}$ does not specify the requested attribute, UA outputs \texttt{unknown}.\footnote{In our implementation, UA is instantiated as an LLM following a strict no-fabrication policy; the prompt is shown in Appendix~\ref{app:prompts} and released with our artifact.}

\paragraph{Rewriter.}
The Rewriter receives only the blurred query $q$ and the clarification Q\&A pairs $\{(c_i,a_i)\}_{i=1}^{k}$, and outputs a single rewritten query $\hat{q}$.
Crucially, the Rewriter must not access $q^{*}$ and must not invent new constraints.
When an answer is \texttt{unknown}, the Rewriter preserves the original ambiguity for that aspect rather than forcing a concrete value.\footnote{The Rewriter prompt is shown in Appendix~\ref{app:prompts} and included in the released artifact.}

\paragraph{Leakage example.}
Consider BQ: ``best universities for AI'' and OQ: ``best universities for AI in Europe for master's study''.
If the Clarifier asks ``Which region are you asking about?'', UA may answer ``Europe'' because this information is explicitly present in OQ.
The Rewriter may then produce ``best universities in Europe for AI''.
However, it cannot add ``for master's study'' unless that constraint is explicitly elicited by a clarification question.
Thus, any added constraint in $\hat{q}$ must be grounded in the observed Q\&A pairs rather than copied from the hidden intent.

This two-phase protocol ensures that downstream improvements are attributable to (i) the clarification questions asked and (ii) how effectively the Rewriter incorporates the elicited answers, rather than recovering intent directly from $q^{*}$.

\subsection{Static Golden Reference}

For each intent query $q^{*}$, we construct a static golden reference by running WebDancer once on $q^{*}$, archiving its tool traces, and extracting a set of evidence-grounded \emph{golden nuggets}.
Each nugget is an atomic, judgeable statement supported by one or more evidence snippets from the archived traces, identified by source IDs.
This yields a stable target that is reused across all Clarifiers and all clarification budgets $k$, enabling reproducible comparison of different clarification strategies.

\section{Dataset}
\label{sec:dataset}

\begin{figure*}[t]
  \centering
  \includegraphics[width=0.95\textwidth]{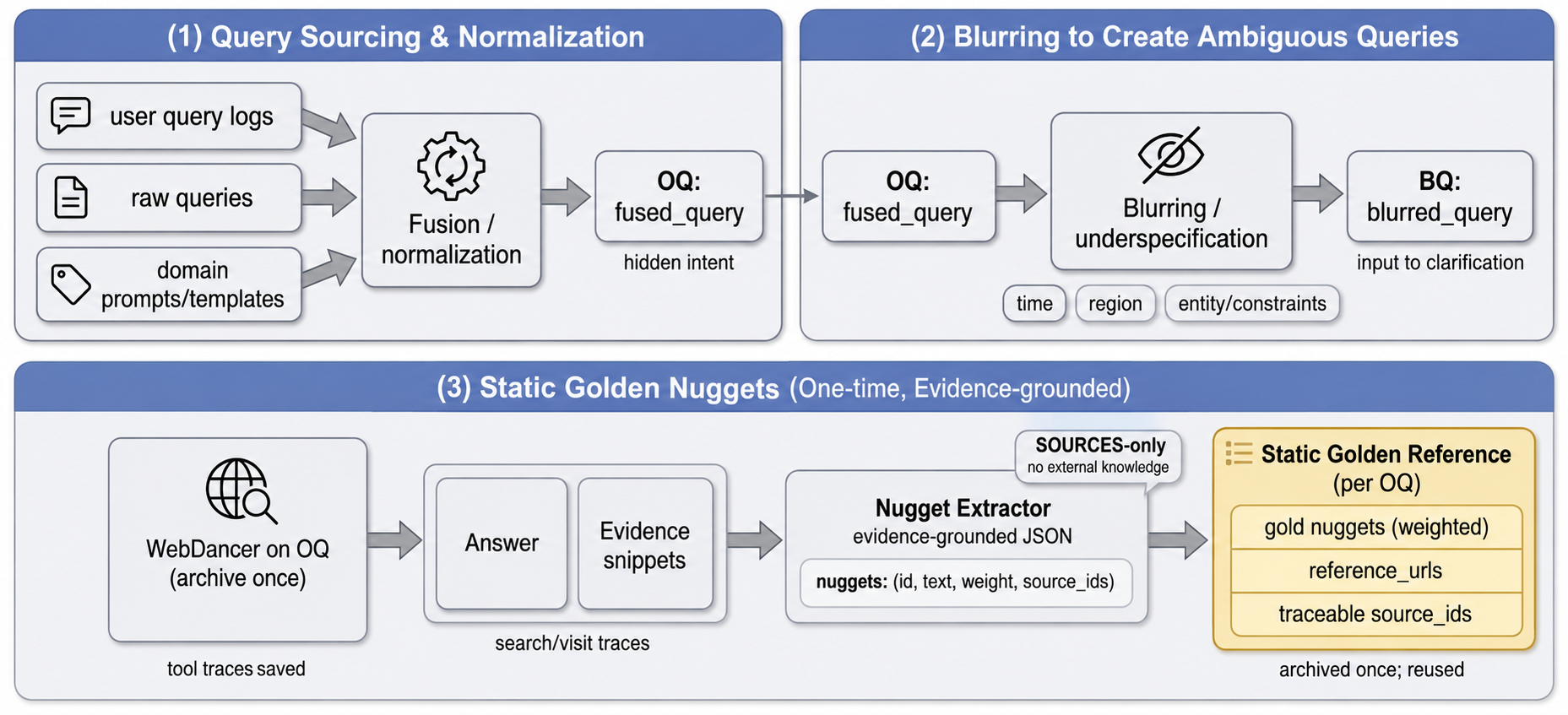}
  \caption{\textbf{Dataset construction and static golden nugget creation.}
  We fuse and normalize raw user queries into intent queries (OQ; \texttt{fused\_query}), then create underspecified benchmark inputs by blurring key constraints to form blurred queries (BQ; \texttt{blurred\_query}).
  For each OQ, we run WebDancer once to archive an answer and search/visit evidence traces, and extract evidence-grounded, weighted golden nuggets with traceable source IDs.
  This static gold is reused to score all systems with \texttt{restore\_score\_100}, enabling reproducible comparisons across models and clarification budgets $k$.}
  \label{fig:dataset_construction}
\end{figure*}

\subsection{Overview and Released Fields}

Our benchmark contains 518 information-seeking queries, deliberately selected to represent underspecified queries where clarification is expected to provide high utility.
Each instance consists of an intent query $q^{*}$ (\texttt{fused\_query}) and an underspecified query $q$ (\texttt{blurred\_query}).
Under the closed-book protocol (Section~\ref{sec:task}), a system observes only $q$, generates $k$ clarification questions, receives constrained answers, rewrites to $\hat{q}$, and is evaluated by running a fixed deep-search backend on $\hat{q}$ and scoring nugget restoration against a static reference built from $q^{*}$.

\paragraph{Why paired blurring?}
Our goal is to construct controlled pairs $(q,q^{*})$ where missing constraints are explicit, so that clarification can be evaluated under a leakage-free protocol.
Rather than synthesizing arbitrary ambiguity, we start from real user intents and remove key constraints such as time, region, entity scope, and criteria.
This paired construction is necessary because $q^{*}$ serves as the hidden intent for UA and as the source query for static gold construction, while $q$ is the only query visible to the Clarifier and Rewriter.

\paragraph{Why we do not release ``gold'' clarification labels.}
Unlike datasets that provide canonical clarification questions as supervised targets, we treat clarification questions as \emph{system outputs}.
Our evaluation asks: \emph{does the model ask for information that is both answerable under the intent and useful for downstream deep search?}
This avoids prescribing a single ``correct'' clarification and instead measures clarification by end-to-end utility.

\paragraph{Released artifacts.}
We release on the project page: (i) all \texttt{fused\_query} and \texttt{blurred\_query} pairs ($n=518$); (ii) a static golden reference per $q^{*}$, including weighted nuggets and traceable evidence identifiers; and (iii) code to reproduce one-time golden construction, Phase2 evaluation runs, and scoring via \texttt{restore\_score\_100}.
Model-generated clarification questions, UA answers, rewrites, and WebDancer outputs are reproducible using the released runner and prompt templates. All data and code are available on the project page.\footnote{\url{https://clarify-then-search.github.io/}}
Figure~\ref{fig:dataset_construction} summarizes dataset construction and the one-time creation of static golden nuggets.

\subsection{Static Golden Nuggets}
\label{sec:gold}

For each intent query $q^{*}$, we construct a static golden reference in two stages.

\paragraph{Stage 1: one-time WebDancer run on $q^{*}$.}
We execute WebDancer once on $q^{*}$ and record (i) the final answer and (ii) tool traces from \texttt{search} and \texttt{visit}.
These traces define the evidence pool used to construct the golden reference and are archived for reproducibility.

\paragraph{Stage 2: evidence-grounded nugget extraction.}
We then apply an evidence-grounded nugget extractor that takes as input $q^{*}$, WebDancer's answer, and the retrieved evidence snippets, and produces a structured \texttt{GOLD} JSON.
The extractor enforces traceability: each nugget is an atomic, judgeable statement and stores identifiers of supporting evidence snippets.
It also records a decomposition of the intent into required sub-questions (\texttt{must\_cover}), missing evidence requirements (\texttt{missing\_evidence}), and unsupported claims that appeared in the answer.
Each nugget is assigned an integer weight $w_j\in\{1,2,3\}$ reflecting its importance for answering $q^{*}$.
The extraction prompt is summarized in Appendix~\ref{app:eval-prompts}, and the full prompt and JSON schema are released with our artifact.

This procedure yields a single archived nugget set per query, which is reused across all evaluated systems and all $k$ settings.
Because the golden reference is built once and held fixed, system comparisons are not affected by run-to-run variation in the deep-search backend during gold construction.

\subsection{Evaluation Inputs and Output Records}

At evaluation time, a system produces clarification questions $C$, UA answers $A$, a rewritten query $\hat{q}$, and a WebDancer output $\hat{y}$ with tool traces.
We compute \texttt{restore\_score\_100} by comparing $\hat{y}$ against the static golden nuggets for $q^{*}$ (Section~\ref{sec:eval}).
We report mean \texttt{restore\_score\_100} as the primary metric, and additionally report distributional summaries including p50, p90, min, and max.

\paragraph{Implementation note.}
In our runner, the candidate answer $\hat{y}$ is extracted from WebDancer's final assistant output by parsing the \texttt{<answer>\ldots</answer>} tag when present, with robust fallbacks for partially closed tags.
All per-instance tool traces are saved for auditing and reproducibility; compact summaries are included in the released evaluation outputs.

\subsection{Evaluation Protocol}
\label{sec:eval}

\subsubsection{Nugget Coverage Judging}

Given a candidate answer $\hat{y}$ and a set of golden nuggets
$\mathcal{N}=\{(n_j,w_j)\}_{j=1}^{m}$,
we determine nugget-level coverage using a strict LLM-based judge.
For each nugget $n_j$, the judge outputs one of three labels:
\texttt{full}, \texttt{partial}, or \texttt{none}.
\texttt{full} indicates that $\hat{y}$ expresses the nugget content with required constraints;
\texttt{partial} indicates that the core statement is present but a critical qualifier is missing;
\texttt{none} indicates that the nugget is absent or contradicted.

We use ERNIE-4.5-Turbo as the coverage judge with deterministic decoding (temperature $0$).
To improve reliability, we apply robust JSON parsing with retries and enforce a strict rubric.
The judge prompt is summarized in Appendix~\ref{app:eval-prompts} and released with our artifact.

\subsubsection{\texttt{restore\_score\_100}}
\label{sec:restore_score}

We define weighted nugget recall with partial credit as
\begin{equation}
\mathrm{Recall}(\hat{y}) =
\frac{\sum_{j=1}^{m} w_j \cdot s(\mathrm{cov}(n_j,\hat{y}))}{\sum_{j=1}^{m} w_j},
\end{equation}
where $\mathrm{cov}(\cdot)\in\{\texttt{full},\texttt{partial},\texttt{none}\}$ is the judge label and
\begin{equation}
s(\texttt{full})=1,\qquad s(\texttt{partial})=0.5,\qquad s(\texttt{none})=0.
\end{equation}
We report
\begin{equation}
\texttt{restore\_score\_100} = 100 \times \mathrm{Recall}(\hat{y}).
\end{equation}
The primary benchmark score is the mean \texttt{restore\_score\_100} over evaluated instances, and we additionally report percentile summaries to characterize distributional behavior.
\section{Experimental Setup}
\label{sec:setup}

\subsection{Phase2 Closed-Book Pipeline and Roles}
\label{sec:pipeline}

We evaluate clarification utility under the controlled closed-book protocol described in Section~\ref{sec:task}.
The pipeline instantiates three LLM roles: a \emph{Clarifier} that generates clarification questions, a \emph{User Answerer (UA)} that answers under a strict no-fabrication policy given the hidden intent, and a \emph{Rewriter} that produces a rewritten query $\hat{q}$ using only the blurred query and the clarification Q\&A pairs.
To isolate clarification ability, we vary only the Clarifier while keeping UA, Rewriter, the deep-search backend, nugget construction, and nugget judging fixed.

Table~\ref{tab:model_backends} lists the model backends used for each pipeline component.
This controlled design attributes performance differences primarily to (i) which ambiguity dimensions are asked about and (ii) whether those questions elicit answerable and useful information under the closed-book UA policy.
The core prompts are shown in Appendices~\ref{app:prompts}--\ref{app:eval-prompts}, and the full executable prompts are released with the benchmark artifact.

\paragraph{Clarifier (varied).}
Given a blurred query $q$ (\texttt{blurred\_query}), the Clarifier outputs $k$ clarification questions $C=(c_1,\ldots,c_k)$, where $k\in\{1,2,3\}$.
Clarifiers are prompted to derive questions from $q$ only, with no access to the intent query $q^{*}$, and to ask questions that target a single ambiguity dimension, such as time, region, entity scope, or definition/criterion.
We enforce strict output formatting as a JSON array of strings and run all Clarifiers with deterministic decoding.

\paragraph{User Answerer (UA; fixed).}
UA is fixed to ERNIE-4.5-Turbo (Table~\ref{tab:model_backends}) with temperature $0$.
For each question $c_i$, UA receives the intent query $q^{*}$ (\texttt{fused\_query}) and must output a minimal answer $a_i$ only if the intent explicitly specifies the requested attribute; otherwise it outputs \texttt{unknown}.
UA is explicitly forbidden from adding any information not present in $q^{*}$.

\paragraph{Rewriter (fixed).}
The Rewriter is also fixed to ERNIE-4.5-Turbo (Table~\ref{tab:model_backends}) with temperature $0$.
It receives only the blurred query $q$ and the clarification Q\&A pairs $\{(c_i,a_i)\}_{i=1}^{k}$, and outputs a single rewritten query $\hat{q}$.
The Rewriter must not introduce new entities or constraints; when $a_i=\texttt{unknown}$, it must preserve the ambiguity rather than forcing a concrete value.

\paragraph{Static gold and judging components (fixed).}
To ensure controlled comparisons across Clarifiers, we also fix the nugget extractor and coverage judge to ERNIE-4.5-Turbo (Table~\ref{tab:model_backends}) with deterministic decoding.
Golden nuggets are extracted from evidence retrieved by running WebDancer once on $q^{*}$ (Section~\ref{sec:gold}).
At evaluation time, nugget coverage is judged with a strict rubric that outputs \texttt{full}/\texttt{partial}/\texttt{none} per nugget (Section~\ref{sec:eval}), with robust JSON parsing and retries.

\begin{table}[t]
\centering
\small
\caption{Model backends used in our benchmark. We vary only the clarification-question generator (Clarifier); UA, Rewriter, nugget extraction, and coverage judging are fixed to ERNIE-4.5-Turbo-128K for controlled comparisons.}
\label{tab:model_backends}
\begin{tabular}{ll}
\toprule
\textbf{Component} & \textbf{Model} \\
\midrule
Clarifier (CQ generator) & Qwen3-235B-A22B-Instruct~\cite{yang2025qwen3} \\
                         & ERNIE-4.5-Turbo-128K~\cite{wang2026ernie} \\
                         & Kimi-K2-Instruct~\cite{team2025kimi} \\
                         & DeepSeek-V3.2~\cite{liu2025deepseek} \\
                         & GPT-5.2~\cite{openai2025gpt52} \\
                         & Claude-Sonnet-4.5~\cite{anthropic2025claude45} \\
                         & Gemini-2.5-Pro~\cite{comanici2025gemini} \\
\midrule
User Answerer (UA) & ERNIE-4.5-Turbo-128K~\cite{wang2026ernie} \\
Rewriter           & ERNIE-4.5-Turbo-128K~\cite{wang2026ernie} \\
Nugget extractor   & ERNIE-4.5-Turbo-128K~\cite{wang2026ernie} \\
Coverage judge     & ERNIE-4.5-Turbo-128K~\cite{wang2026ernie} \\
\bottomrule
\end{tabular}
\end{table}

\subsection{Deep Search Backend: WebDancer}
\label{sec:webdancer}

We use WebDancer~\cite{wu2025webdancer} as the fixed deep-search backend for all evaluations.
Given a rewritten query $\hat{q}$, WebDancer interleaves reasoning with tool calls to \texttt{search} and \texttt{visit}, and produces a final answer $\hat{y}$ along with tool traces.
We run WebDancer in batch mode with a fixed tool budget and fixed configuration, and persist per-instance tool traces for auditing and reproducibility.
The candidate answer $\hat{y}$ is extracted from WebDancer's final output by parsing the \texttt{<answer>\ldots</answer>} tag when present, with robust fallbacks for partially closed tags.

Fixing WebDancer is a deliberate choice for controlled evaluation.
Our goal is not to claim backend-independent absolute scores, but to compare Clarifiers under the same deep-search environment.
This avoids conflating differences in clarification quality with variation from different retrieval or agentic search backends.

\subsection{Evaluation Set and Reporting}
\label{sec:eval_sets}

\paragraph{Evaluation set.}
All experiments and analyses in this paper are conducted on our released dataset of 518 instances (Section~\ref{sec:dataset}).
These instances are selected to represent underspecified queries where clarification is expected to matter most, so we do not further partition them into difficulty subsets in the main paper.

\paragraph{Budgeted clarification protocol.}
For each query and each budget $k\in\{1,2,3\}$, we execute:
(1) the Clarifier generates $k$ questions $C$;
(2) UA generates $k$ answers $A$ under the closed-book constraint;
(3) the Rewriter outputs $\hat{q}$ using only $q$ and $\{(c_i,a_i)\}_{i=1}^{k}$;
and (4) WebDancer runs on $\hat{q}$ to produce a candidate answer $\hat{y}$.
We then compute \texttt{restore\_score\_100} of $\hat{y}$ against the static golden nuggets archived from a one-time WebDancer run on $q^{*}$ (Section~\ref{sec:gold}).

\paragraph{Metrics and reported statistics.}
We report mean \texttt{restore\_score\_100} as the primary score, and additionally report p50/p90 and min/max to characterize distributional behavior.
All prompts, scripts, and configurations required to reproduce the full pipeline are released with the benchmark.
\section{Results}
\label{sec:results}

\subsection{Main Results: One-Turn Clarification}
\label{sec:results_k1}

We first evaluate clarification utility under the closed-book Phase2 protocol with a single clarification question ($k=1$) on the 518-instance benchmark.
Table~\ref{tab:hard_k1_main} reports \texttt{restore\_score\_100} for each Clarifier, where \texttt{orig} denotes running WebDancer directly on the underspecified query without clarification.

All Clarifiers improve over \texttt{orig}, showing that even one clarification can elicit useful intent-grounded information for rewriting and downstream deep search.
However, the magnitude of improvement varies across models.
Among the evaluated open-weight models, including Qwen, ERNIE, DeepSeek, and Kimi, mean gains are modest (+3.09 to +4.03), and the median remains close to the baseline, indicating that one-turn clarification often fails to recover decisive constraints.
In contrast, GPT, Claude and Gemini yield larger gains (+6.44 to +7.04), with GPT achieving the highest mean and median.
Beyond the mean, clarification substantially lifts the recoverable tail: p90 increases from 35.3 for \texttt{orig} to roughly 46--53 across Clarifiers, suggesting that some instances become much more recoverable once a key ambiguity is resolved.

\paragraph{Statistical robustness.}
We compute paired bootstrap 95\% confidence intervals (CIs) on per-instance score differences against \texttt{orig}.
As shown in Table~\ref{tab:hard_k1_ci}, all CIs are strictly above zero, confirming that the improvements are not driven by a small number of outliers.

\begin{table}[t]
\centering
\small
\caption{Benchmark results ($n=518$) under closed-book Phase2 with $k=1$. Metric: \texttt{restore\_score\_100}. $\Delta$ is mean difference vs.\ \texttt{orig}.}
\label{tab:hard_k1_main}
\begin{tabular}{lccc}
\toprule
System & mean ($\Delta$) & p50 & p90 \\
\midrule
orig      & 19.434 & 20.454 & 35.294 \\
Qwen      & 22.527 (+3.093) & 19.565 & 47.211 \\
ERNIE        & 23.460 (+4.026) & 20.000 & 50.000 \\
DeepSeek  & 23.234 (+3.799) & 21.053 & 48.416 \\
Kimi      & 22.687 (+3.252) & 20.000 & 46.236 \\
GPT       & \textbf{26.474} (\textbf{+7.040}) & \textbf{25.000} & 50.000 \\
Claude    & 25.911 (+6.477) & 23.333 & \textbf{52.996} \\
Gemini    & 25.874 (+6.440) & 21.429 & \textbf{52.996} \\
\bottomrule
\end{tabular}
\end{table}

\begin{table}[t]
\centering
\small
\caption{Mean improvement over \texttt{orig} at $k=1$ with paired bootstrap 95\% CIs.}
\label{tab:hard_k1_ci}
\begin{tabular}{lcc}
\toprule
System & mean $\Delta$ vs.\ \texttt{orig} & 95\% CI \\
\midrule
Qwen     & +3.093 & [1.482, 4.729] \\
ERNIE    & +4.026 & [2.526, 5.637] \\
DeepSeek & +3.799 & [2.338, 5.357] \\
Kimi     & +3.252 & [1.753, 4.762] \\
GPT      & \textbf{+7.040} & [5.566, 8.578] \\
Claude   & +6.477 & [4.890, 8.159] \\
Gemini   & +6.440 & [4.902, 8.057] \\
\bottomrule
\end{tabular}
\end{table}

\subsection{UA Diagnostics: Unknown Rate and Question-Type Bias}
\label{sec:results_ua_diag}

Phase2 exposes an intermediate information bottleneck: a clarification question is useful only if it elicits an intent-grounded answer that can be safely incorporated into the rewritten query.
We therefore analyze UA responses and question types on the 518 instances.

For each clarification question, UA returns \texttt{unknown} when the hidden intent does not explicitly specify the requested attribute.
At $k=1$, \texttt{ua\_unknown\_rate} is also the probability that the interaction yields no grounded signal.
Table~\ref{tab:ua_overall_k1} shows that \texttt{ua\_unknown\_rate} remains high across Clarifiers (0.600--0.701).
Thus, in roughly two thirds of instances, one clarification question fails to retrieve any intent-grounded constraint, limiting what the Rewriter can safely add and leaving the rewritten query close to the original underspecified input.

A consistent low-yield failure mode is \emph{region-only} clarification, such as asking for a city or province.
Many intents do not specify a location, so such questions frequently elicit \texttt{unknown}.
Table~\ref{tab:ua_regiononly_k1} shows that region-only questions have very high unknown rates (0.750--0.869), while models differ substantially in how often they ask them: \texttt{cq\_region\_rate} ranges from 0.143 for Kimi to 0.498 for Gemini.
This illustrates why plausible clarification is not necessarily useful clarification: a question can be natural and topical, yet yield no usable information under the closed-book constraint.

\begin{table}[t]
\centering
\small
\caption{UA diagnostics at $k=1$ ($n=518$). \texttt{ua\_unknown\_rate}: fraction of answers labeled \texttt{unknown}. \texttt{cq\_region\_rate}: fraction of clarification questions about region/location.}
\label{tab:ua_overall_k1}
\begin{tabular}{lcc}
\toprule
Model & ua\_unknown\_rate & cq\_region\_rate \\
\midrule
Qwen     & 0.620 & 0.195 \\
ERNIE    & 0.680 & 0.357 \\
DeepSeek & 0.622 & 0.479 \\
Kimi     & 0.666 & 0.143 \\
GPT      & 0.618 & 0.351 \\
Claude   & 0.600 & 0.326 \\
Gemini   & 0.701 & 0.498 \\
\bottomrule
\end{tabular}
\end{table}

\begin{table}[t]
\centering
\small
\caption{Region-only clarification as a low-yield failure mode at $k=1$ ($n=518$). We report the number of region-only questions and the UA unknown rate conditioned on region-only.}
\label{tab:ua_regiononly_k1}
\begin{tabular}{lcc}
\toprule
Model & \#region-only & region-only unknown \\
\midrule
Qwen     &  64 & 0.766 \\
ERNIE       & 136 & 0.787 \\
DeepSeek & 223 & 0.807 \\
Kimi     &  48 & 0.750 \\
GPT      & 121 & 0.785 \\
Claude   & 148 & 0.824 \\
Gemini   & 236 & 0.869 \\
\bottomrule
\end{tabular}
\end{table}

\subsection{Budgeted Clarification: $k=2$ and $k=3$}
\label{sec:results_k23}

A single-turn interaction may miss the most useful ambiguity dimension.
We therefore evaluate larger clarification budgets $k\in\{2,3\}$ under the same closed-book protocol.
Table~\ref{tab:hard_k23} summarizes the results.

Increasing the budget from $k=1$ to $k=2$ yields a clear jump in mean restoration for all models, with gains ranging from +6.16 to +7.64 over \texttt{orig}.
At $k=3$, ERNIE achieves the largest improvement (+8.90), while some models show saturation or slight regression relative to $k=2$.
The median also increases with budget, indicating that additional turns make improvements more typical rather than only improving a small tail.

The relative ordering changes with larger budgets.
At $k = 1$, GPT achieves the highest mean score, while ERNIE achieves the highest mean among all open-weight models. At $k = 3$, ERNIE becomes the top-performing model. This suggests that larger budgets reward sustained question-selection quality across turns, rather than a single high-impact question.

\paragraph{Multi-turn unknown rate.}
We further analyze how often an entire interaction yields no grounded signal.
Table~\ref{tab:ua_budget_unknown} reports \texttt{all\_unknown\_rate} and the number of non-\texttt{unknown} answers per instance.
Increasing $k$ substantially reduces \texttt{all\_unknown\_rate}: for example, GPT drops from 0.618 at $k=1$ to 0.375 at $k=2$ and 0.189 at $k=3$.
This aligns with the restoration gains, suggesting that larger budgets help mainly by reducing no-signal interactions and increasing the chance of eliciting at least one useful constraint.

\begin{table}[t]
\centering
\small
\caption{Benchmark results under Phase2 for $k\in\{2,3\}$ ($n=518$). Metric: \texttt{restore\_score\_100}. $\Delta$ is mean difference vs.\ \texttt{orig}.}
\label{tab:hard_k23}
\begin{tabular}{llccc}
\toprule
$k$ & System & mean ($\Delta$) & p50 & p90 \\
\midrule
\multirow{8}{*}{$2$}
& orig     & 19.434 & 20.454 & 35.294 \\
& Qwen     & 26.020 (+6.586) & 23.607 & 52.941 \\
& ERNIE       & 26.351 (+6.917) & 23.842 & 52.632 \\
& DeepSeek & 26.628 (+7.194) & 23.747 & 53.259 \\
& Kimi     & 25.593 (+6.159) & 22.902 & 50.000 \\
& GPT      & \textbf{27.073} (\textbf{+7.639}) & \textbf{25.000} & 53.487 \\
& Claude   & 26.203 (+6.769) & 23.509 & 53.654 \\
& Gemini   & 26.904 (+7.470) & \textbf{25.000} & \textbf{54.545} \\
\midrule
\multirow{8}{*}{$3$}
& orig     & 19.434 & 20.454 & 35.294 \\
& Qwen     & 26.235 (+6.801) & 22.997 & 51.744 \\
& ERNIE       & \textbf{28.339} (\textbf{+8.904}) & \textbf{26.087} & \textbf{57.143} \\
& DeepSeek & 26.486 (+7.052) & 25.000 & 52.424 \\
& Kimi     & 26.568 (+7.134) & 23.509 & 56.310 \\
& GPT      & 26.297 (+6.863) & 23.385 & 54.407 \\
& Claude   & 27.029 (+7.594) & 25.000 & 53.329 \\
& Gemini   & 27.126 (+7.691) & 24.194 & 52.628 \\
\bottomrule
\end{tabular}
\end{table}

\begin{table}[t]
\centering
\small
\caption{Interaction-level UA unknown diagnostics. \texttt{all\_unknown}: all $k$ answers are \texttt{unknown}. \texttt{known\_cnt}: number of non-\texttt{unknown} answers.}
\label{tab:ua_budget_unknown}
\begin{tabular}{lcccc}
\toprule
$k$ / Model & all\_unknown & known\_cnt=1 & known\_cnt=2 & known\_cnt=3 \\
\midrule
\multicolumn{5}{c}{$k=1$} \\
Qwen     & 0.620 & 0.380 & -- & -- \\
ERNIE       & 0.680 & 0.320 & -- & -- \\
DeepSeek & 0.622 & 0.378 & -- & -- \\
Kimi     & 0.666 & 0.334 & -- & -- \\
GPT      & 0.618 & 0.382 & -- & -- \\
Claude   & 0.600 & 0.400 & -- & -- \\
Gemini   & 0.701 & 0.299 & -- & -- \\
\midrule
\multicolumn{5}{c}{$k=2$} \\
Qwen     & 0.446 & 0.398 & 0.156 & -- \\
ERNIE       & 0.498 & 0.390 & 0.112 & -- \\
DeepSeek & 0.394 & 0.452 & 0.154 & -- \\
Kimi     & 0.510 & 0.407 & 0.083 & -- \\
GPT      & \textbf{0.375} & 0.458 & 0.168 & -- \\
Claude   & 0.452 & 0.409 & 0.139 & -- \\
Gemini   & 0.523 & 0.380 & 0.097 & -- \\
\midrule
\multicolumn{5}{c}{$k=3$} \\
Qwen     & 0.417 & 0.417 & 0.139 & 0.027 \\
ERNIE       & 0.394 & 0.403 & 0.164 & 0.039 \\
DeepSeek & 0.251 & 0.396 & 0.270 & 0.083 \\
Kimi     & 0.421 & 0.427 & 0.137 & 0.015 \\
GPT      & \textbf{0.189} & 0.432 & 0.315 & 0.064 \\
Claude   & 0.295 & 0.438 & 0.226 & 0.041 \\
Gemini   & 0.413 & 0.398 & 0.164 & 0.025 \\
\bottomrule
\end{tabular}
\end{table}

\section{Analysis}
\label{sec:analysis}

\subsection{Information Gain is the Main Bottleneck}
\label{sec:analysis_infogain}

A key advantage of our benchmark is that it makes intermediate interaction behavior measurable, rather than judging only the final deep-search answer.
Under the closed-book protocol, we can inspect both what clarification questions a model asks and whether the hidden intent supports a grounded answer.
This allows us to distinguish two factors that are often conflated in interactive search: whether a question sounds reasonable, and whether it actually yields usable information for rewriting.

The results show that information gain is often the bottleneck.
At $k=1$, UA \texttt{unknown} rates remain high across models (0.600--0.701; Table~\ref{tab:ua_overall_k1}), meaning that many single-turn clarifications elicit no intent-grounded constraint.
When the only answer is \texttt{unknown}, the Rewriter has little grounded content to incorporate and must proceed with residual ambiguity.
This helps explain why one-turn clarification reliably improves over \texttt{orig} but still leaves substantial room for improvement.

Multi-turn statistics further support this interpretation.
As shown in Table~\ref{tab:ua_budget_unknown}, increasing the clarification budget reduces the probability that an entire interaction yields no grounded signal.
For example, GPT's \texttt{all\_unknown} rate drops from 0.618 at $k=1$ to 0.375 at $k=2$ and 0.189 at $k=3$.
This aligns with the restoration gains in Table~\ref{tab:hard_k23}: additional turns help mainly by increasing the chance of eliciting at least one usable constraint.
However, the marginal gain depends on whether later questions target orthogonal and answerable ambiguity dimensions, rather than repeatedly asking about attributes absent from the intent.

\subsection{Question-Type Bias is a Measurable Failure Mode}
\label{sec:analysis_qtype}

Our diagnostics reveal a concrete failure mode: plausible-sounding clarification is not necessarily useful clarification.
A prominent example is region-only questioning.
At $k=1$, models differ substantially in their tendency to ask region/location questions: \texttt{cq\_region\_rate} ranges from 0.143 for Kimi to 0.498 for Gemini (Table~\ref{tab:ua_overall_k1}).
However, region-only questions are consistently low-yield under our closed-book protocol.
As shown in Table~\ref{tab:ua_regiononly_k1}, their UA unknown rates are very high across all models (0.750--0.869).

This low-yield pattern persists beyond one-turn clarification.
Table~\ref{tab:region_summary} summarizes the full turn-level diagnostics, with the complete
turn-level region-only statistics provided in Appendix~\ref{app:additional-diagnostics}: across larger
budgets, region-only unknown rates remain high across models and turns.
This indicates that repeatedly asking for location does not reliably increase information gain when the hidden intent does not contain location information.

\begin{table}[t]
\centering
\small
\caption{Region-only clarification remains low-yield across clarification budgets. We report the range of UA \texttt{unknown} rates for region-only questions across models and turns.}
\label{tab:region_summary}
\begin{tabular}{lcc}
\toprule
Setting & Region-only unknown range & Observation \\
\midrule
$k=1$ & 0.750--0.869 & consistently high \\
$k=2$ & 0.800--0.917 & remains high across turns \\
$k=3$ & 0.692--1.000 & often still unanswerable \\
\bottomrule
\end{tabular}
\end{table}

This pattern matters because a region question can appear natural and relevant even when the hidden intent does not specify any location.
In such cases, the answer becomes \texttt{unknown}, and the Rewriter is not allowed to add a concrete region.
Thus, region bias can waste clarification budget without improving the rewritten query.
By exposing this behavior, the benchmark supports diagnostics beyond surface fluency or topical relevance and identifies which question types are likely to produce low information gain.

\subsection{Budgeted Clarification as a Controlled Intervention}
\label{sec:analysis_k}

Varying the clarification budget $k$ provides a controlled way to test whether performance is limited by one-shot question selection or by deeper limitations of the pipeline.
The restoration results show a clear budget effect: moving from $k=1$ to $k=2$ increases mean improvements for all models, and $k=3$ further improves several systems (Table~\ref{tab:hard_k23}).
At the same time, the gains are not purely monotonic for every model, suggesting that extra turns are useful only when they elicit additional grounded constraints.

The relative ordering also changes with budget.
GPT performs best at $k=1$ and $k=2$, while ERNIE becomes strongest at $k=3$.
This suggests that different budgets reward different clarification abilities: $k=1$ emphasizes single-shot targeting of a high-value ambiguity, whereas larger budgets reward sustained question-selection quality across turns.
Therefore, evaluating only one clarification budget can be misleading for understanding model behavior.

These observations also motivate an ask-or-not decision.
A practical clarify-then-search system should ask only when the expected information gain is positive.
Signals such as \texttt{all\_unknown} rate and question-type-specific unknown rate provide measurable proxies for learning or evaluating such gating policies.
For example, high predicted answerability may justify asking, while low-yield question types such as region-only clarification may suggest abstaining or switching to retrieval-first exploration.

\subsection{Robustness to Supporting and Judging Backbones}
\label{sec:analysis_robustness}

Our main experiments fix UA, Rewriter, nugget extraction, and coverage judging to ensure controlled comparisons.
To examine whether the main conclusions depend on a single ERNIE-based stack, we further conduct a backbone-substitution analysis for the supporting components and coverage judge.

For the end-to-end judge, replacing the ERNIE-based coverage judge changes the absolute score scale, but preserves the main ordering.
As shown in Table~\ref{tab:judge_robustness}, under ERNIE-, GPT-, and Claude-based judges, both ERNIE-$k1$ and GPT-$k1$ outperform \texttt{orig}, and the ordering GPT $>$ ERNIE $>$ \texttt{orig} remains stable.
This suggests that the main end-to-end conclusion is not an artifact of one particular coverage judge.

For the intermediate stages, UA shows moderate backbone sensitivity, mainly in the overall \texttt{unknown} rate and answer style.
However, pairwise known/unknown agreement remains relatively high across supporting backbones, indicating that different UA backbones often agree at the coarse answerability level.
In contrast, Rewriter protocol behavior is highly stable: unknown-preservation rates remain above 0.98, and unsupported concretization, such as adding a new year or region when the answer is \texttt{unknown}, is near zero.
Overall, these results support our design choice of fixing supporting components
for controlled comparison while showing that the primary findings are robust
to reasonable backbone substitutions.
Detailed supporting-backbone stability statistics are provided in
Appendix~\ref{app:additional-diagnostics}.

\begin{table}[t]
\centering
\small
\caption{Judge-backbone robustness at $k=1$ on the benchmark set. Absolute scores vary across judges, but the improvement over \texttt{orig} and the relative ordering remain stable.}
\label{tab:judge_robustness}
\begin{tabular}{lcccc}
\toprule
Judge & \texttt{orig} & ERNIE-$k1$ & GPT-$k1$ & Ordering \\
\midrule
ERNIE  & 19.43 & 23.46 & 26.47 & GPT $>$ ERNIE $>$ orig \\
GPT    & 27.74 & 29.54 & 29.79 & GPT $>$ ERNIE $>$ orig \\
Claude & 32.19 & 33.80 & 34.62 & GPT $>$ ERNIE $>$ orig \\
\bottomrule
\end{tabular}
\end{table}

\section{Conclusion}
We introduced Clarify-Then-Search, a benchmark for evaluating whether clarification improves deep search under realistic information constraints. The benchmark uses a two-phase closed-book protocol and static evidence-grounded golden nuggets to measure end-to-end nugget restoration after clarification, rewriting, and deep search. Our experiments show that clarification consistently improves over the no-interaction baseline, and that larger clarification budgets generally yield stronger restoration, with ERNIE-4.5-Turbo-128K becoming the overall top-performing model at $k=3$. At the same time, diagnostic analyses reveal that many clarification turns still elicit unknown or low-yield answers, especially for region-only questions, highlighting the need for future systems to better decide when to ask, what to ask, and how to use clarification signals for robust deep search.

\section*{Limitations}
First, our closed-book protocol uses an LLM-based UA to produce constrained answers from the hidden intent query.
Although this design prevents intent leakage to the Rewriter, the exact \texttt{unknown} policy can affect measured utility.
Second, our end-to-end evaluation fixes WebDancer as the deep-search backend.
The benchmark is therefore best interpreted as a controlled evaluation of clarification utility under a shared deep-search pipeline, rather than a backend-agnostic ranking of all possible search systems.
Third, golden nugget construction and coverage judging rely on LLMs with strict rubrics and deterministic decoding; borderline semantic matches may still introduce judging noise.

\begin{acks}
This research was supported in part by the National Natural Science Foundation of China under Grant No.\ 92370204, the National Science and Technology Major Project of China under Grant No.\ 2023ZD0121104, and the Anhui Natural Science Foundation under Grant No.\ 2508085ZD006.
\end{acks}
\clearpage
\balance

\bibliographystyle{ACM-Reference-Format}
\bibliography{reference}

@inproceedings{Aliannejadi2019Clarify,
  title={Asking clarifying questions in open-domain information-seeking conversations},
  author={Aliannejadi, Mohammad and Zamani, Hamed and Crestani, Fabio and Croft, W Bruce},
  booktitle={Proceedings of the 42nd international acm sigir conference on research and development in information retrieval},
  pages={475--484},
  year={2019}
}

@inproceedings{Zamani2020Mimics,
title={Mimics: A large-scale data collection for search clarification},
  author={Zamani, Hamed and Lueck, Gord and Chen, Everest and Quispe, Rodolfo and Luu, Flint and Craswell, Nick},
  booktitle={Proceedings of the 29th ACM international conference on information \& knowledge management},
  pages={3189--3196},
  year={2020}
}

@inproceedings{Zamani2020ClarifyIR,
title={Generating clarifying questions for information retrieval},
  author={Zamani, Hamed and Dumais, Susan and Craswell, Nick and Bennett, Paul and Lueck, Gord},
  booktitle={Proceedings of the web conference 2020},
  pages={418--428},
  year={2020}
}

@article{Dalton2020CAsT19,
  title={TREC CAsT 2019: The conversational assistance track overview},
  author={Dalton, Jeffrey and Xiong, Chenyan and Callan, Jamie},
  journal={arXiv preprint arXiv:2003.13624},
  year={2020}
}

@inproceedings{wu-etal-2022-conqrr,
  title={Conqrr: Conversational query rewriting for retrieval with reinforcement learning},
  author={Wu, Zeqiu and Luan, Yi and Rashkin, Hannah and Reitter, David and Hajishirzi, Hannaneh and Ostendorf, Mari and Tomar, Gaurav Singh},
  booktitle={Proceedings of the 2022 Conference on Empirical Methods in Natural Language Processing},
  pages={10000--10014},
  year={2022}
}

@inproceedings{ye-etal-2023-enhancing,
  title={Enhancing conversational search: Large language model-aided informative query rewriting},
  author={Ye, Fanghua and Fang, Meng and Li, Shenghui and Yilmaz, Emine},
  journal={arXiv preprint arXiv:2310.09716},
  year={2023}
}

@article{Wei2025BrowseComp,
  title={Browsecomp: A simple yet challenging benchmark for browsing agents},
  author={Wei, Jason and Sun, Zhiqing and Papay, Spencer and McKinney, Scott and Han, Jeffrey and Fulford, Isa and Chung, Hyung Won and Passos, Alex Tachard and Fedus, William and Glaese, Amelia},
  journal={arXiv preprint arXiv:2504.12516},
  year={2025}
}

@inproceedings{Li2023APIBank,
  title={Api-bank: A comprehensive benchmark for tool-augmented llms},
  author={Li, Minghao and Zhao, Yingxiu and Yu, Bowen and Song, Feifan and Li, Hangyu and Yu, Haiyang and Li, Zhoujun and Huang, Fei and Li, Yongbin},
  journal={arXiv preprint arXiv:2304.08244},
  year={2023}
}

@inproceedings{Yao2023ReAct,
  title={React: Synergizing reasoning and acting in language models},
  author={Yao, Shunyu and Zhao, Jeffrey and Yu, Dian and Du, Nan and Shafran, Izhak and Narasimhan, Karthik R and Cao, Yuan},
  booktitle={The eleventh international conference on learning representations},
  year={2022}
}

@inproceedings{Dang2007Pyramid,
  title={Different structures for evaluating answers to complex questions: Pyramids won’t topple, and neither will human assessors},
  author={Dang, Hoa Trang and Lin, Jimmy},
  booktitle={Proceedings of the 45th Annual Meeting of the Association of Computational Linguistics},
  pages={768--775},
  year={2007}
}

@article{Lin2006Pourpre,
  title={Methods for automatically evaluating answers to complex questions},
  author={Lin, Jimmy and Demner-Fushman, Dina},
  journal={Information Retrieval},
  volume={9},
  number={5},
  pages={565--587},
  year={2006},
  publisher={Springer}
}

@article{Pradeep2024AutoNugget,
  title={Initial nugget evaluation results for the trec 2024 rag track with the autonuggetizer framework},
  author={Pradeep, Ronak and Thakur, Nandan and Upadhyay, Shivani and Campos, Daniel and Craswell, Nick and Lin, Jimmy},
  journal={arXiv preprint arXiv:2411.09607},
  year={2024}
}

@inproceedings{Aliannejadi2020ClariQ,
  title={ClarQ: A large-scale and diverse dataset for clarification question generation},
  author={Kumar, Vaibhav and Black, Alan W},
  booktitle={Proceedings of the 58th annual meeting of the association for computational linguistics},
  pages={7296--7301},
  year={2020}
}

@article{Tavakoli2022MIMICSDuo,
  title={Mimics-duo: Offline \& online evaluation of search clarification},
  author={Tavakoli, Leila and Trippas, Johanne R and Zamani, Hamed and Scholer, Falk and Sanderson, Mark},
  booktitle={Proceedings of the 45th International ACM SIGIR Conference on Research and Development in Information Retrieval},
  pages={3198--3208},
  year={2022}
}

@inproceedings{Bi2021CQNegative,
  title={Asking clarifying questions based on negative feedback in conversational search},
  author={Bi, Keping and Ai, Qingyao and Croft, W Bruce},
  booktitle={Proceedings of the 2021 ACM SIGIR International Conference on Theory of Information Retrieval},
  pages={157--166},
  year={2021}
}

@article{Elgohary2019CANARD,
  title={Can you unpack that? learning to rewrite questions-in-context},
  author={Elgohary, Ahmed and Peskov, Denis and Boyd-Graber, Jordan},
  journal={Can You Unpack That? Learning to Rewrite Questions-in-Context},
  year={2019}
}

@inproceedings{Anantha2021QReCC,
  title={Open-domain question answering goes conversational via question rewriting},
  author={Anantha, Raviteja and Vakulenko, Svitlana and Tu, Zhucheng and Longpre, Shayne and Pulman, Stephen and Chappidi, Srinivas},
  booktitle={Proceedings of the 2021 Conference of the North American Chapter of the Association for Computational Linguistics: Human Language Technologies},
  pages={520--534},
  year={2021}
}

@inproceedings{Zhou2023WebArena,
  title={Webarena: A realistic web environment for building autonomous agents},
  author={Zhou, Shuyan and Xu, Frank F and Zhu, Hao and Zhou, Xuhui and Lo, Robert and Sridhar, Abishek and Cheng, Xianyi and Ou, Tianyue and Bisk, Yonatan and Fried, Daniel and others},
  booktitle={International Conference on Learning Representations},
  volume={2024},
  pages={15585--15606},
  year={2024}
}

@article{Yao2022WebShop,
  title={Webshop: Towards scalable real-world web interaction with grounded language agents},
  author={Yao, Shunyu and Chen, Howard and Yang, John and Narasimhan, Karthik},
  journal={Advances in Neural Information Processing Systems},
  volume={35},
  pages={20744--20757},
  year={2022}
}

@article{Liu2023AgentBench,
  title={Agentbench: Evaluating llms as agents},
  author={Liu, Xiao and Yu, Hao and Zhang, Hanchen and Xu, Yifan and Lei, Xuanyu and Lai, Hanyu and Gu, Yu and Ding, Hangliang and Men, Kaiwen and Yang, Kejuan and others},
  journal={arXiv preprint arXiv:2308.03688},
  year={2023}
}

@article{Qin2023ToolBench,
  title={Toolllm: Facilitating large language models to master 16000+ real-world apis},
  author={Qin, Yujia and Liang, Shihao and Ye, Yining and Zhu, Kunlun and Yan, Lan and Lu, Yaxi and Lin, Yankai and Cong, Xin and Tang, Xiangru and Qian, Bill and others},
  journal={arXiv preprint arXiv:2307.16789},
  year={2023}
}

@inproceedings{Voorhees2004TRECQA,
  title={Overview of the TREC 2003 Question Answering Track.},
  author={Voorhees, Ellen M and Buckland, L},
  booktitle={TREC},
  volume={2003},
  pages={54--68},
  year={2003}
}

@article{yang2025qwen3,
  title={Qwen3 technical report},
  author={Yang, An and Li, Anfeng and Yang, Baosong and Zhang, Beichen and Hui, Binyuan and Zheng, Bo and Yu, Bowen and Gao, Chang and Huang, Chengen and Lv, Chenxu and others},
  journal={arXiv preprint arXiv:2505.09388},
  year={2025}
}

@article{team2025kimi,
  title={Kimi k2: Open agentic intelligence},
  author={Team, Kimi and Bai, Yifan and Bao, Yiping and Chen, Guanduo and Chen, Jiahao and Chen, Ningxin and Chen, Ruijue and Chen, Yanru and Chen, Yuankun and Chen, Yutian and others},
  journal={arXiv preprint arXiv:2507.20534},
  year={2025}
}

@article{liu2025deepseek,
  title={Deepseek-v3. 2: Pushing the frontier of open large language models},
  author={Liu, Aixin and Mei, Aoxue and Lin, Bangcai and Xue, Bing and Wang, Bingxuan and Xu, Bingzheng and Wu, Bochao and Zhang, Bowei and Lin, Chaofan and Dong, Chen and others},
  journal={arXiv preprint arXiv:2512.02556},
  year={2025}
}

@article{comanici2025gemini,
  title={Gemini 2.5: Pushing the frontier with advanced reasoning, multimodality, long context, and next generation agentic capabilities},
  author={Comanici, Gheorghe and Bieber, Eric and Schaekermann, Mike and Pasupat, Ice and Sachdeva, Noveen and Dhillon, Inderjit and Blistein, Marcel and Ram, Ori and Zhang, Dan and Rosen, Evan and others},
  journal={arXiv preprint arXiv:2507.06261},
  year={2025}
}

@misc{anthropic2025claude45,
  title        = {Introducing Claude Sonnet 4.5},
  author       = {{Anthropic}},
  year         = {2025},
  howpublished = {\url{https://www.anthropic.com/news/claude-sonnet-4-5}},
  note         = {Accessed: 2026-02-07}
}

@misc{openai2025gpt52,
  title        = {Introducing GPT-5.2},
  author       = {{OpenAI}},
  year         = {2025},
  howpublished = {\url{https://openai.com/index/introducing-gpt-5-2/}},
  note         = {Accessed: 2026-02-07}
}

@article{wang2026ernie,
  title={ERNIE 5.0 Technical Report},
  author={Wang, Haifeng and Wu, Hua and Wu, Tian and Sun, Yu and Liu, Jing and Yu, Dianhai and Ma, Yanjun and He, Jingzhou and He, Zhongjun and Hong, Dou and others},
  journal={arXiv preprint arXiv:2602.04705},
  year={2026}
}

@inproceedings{wu2025webdancer,
  title={Webdancer: Towards autonomous information seeking agency},
  author={Wu, Jialong and Li, Baixuan and Fang, Runnan and Yin, Wenbiao and Zhang, Liwen and Wang, Zhenglin and Tao, Zhengwei and Zhang, Ding-Chu and Xi, Zekun and Tang, Robert and others},
  journal={Advances in Neural Information Processing Systems},
  volume={38},
  pages={120957--120985},
  year={2026}
}

@inproceedings{huang-etal-2025-manusearch,
  title={ManuSearch: Democratizing Deep Search in Large Language Models with a Transparent and Open Multi-Agent Framework},
  author={Huang, Lisheng and Liu, Yichen and Jiang, Jinhao and Zhang, Rongxiang and Yan, Jiahao and Li, Junyi and Zhao, Wayne Xin},
  booktitle={Findings of the Association for Computational Linguistics: EMNLP 2025},
  pages={2403--2417},
  year={2025}
}

@inproceedings{luo2025imagescope,
  title={Imagescope: Unifying language-guided image retrieval via large multimodal model collective reasoning},
  author={Luo, Pengfei and Zhou, Jingbo and Xu, Tong and Xia, Yuan and Xu, Linli and Chen, Enhong},
  booktitle={Proceedings of the ACM on Web Conference 2025},
  pages={1666--1682},
  year={2025}
}

@inproceedings{he2024webvoyager,
  title={Webvoyager: Building an end-to-end web agent with large multimodal models},
  author={He, Hongliang and Yao, Wenlin and Ma, Kaixin and Yu, Wenhao and Dai, Yong and Zhang, Hongming and Lan, Zhenzhong and Yu, Dong},
  booktitle={Proceedings of the 62nd Annual Meeting of the Association for Computational Linguistics (Volume 1: Long Papers)},
  pages={6864--6890},
  year={2024}
}

@article{zhang2026omegause,
  title={OmegaUse: Building a General-Purpose GUI Agent for Autonomous Task Execution},
  author={Zhang, Le and Xiao, Yixiong and Lu, Xinjiang and Cao, Jingjia and Zhao, Yusai and Zhou, Jingbo and An, Lang and Feng, Zikan and Sha, Wanxiang and Shi, Yu and others},
  journal={arXiv preprint arXiv:2601.20380},
  year={2026}
}

@inproceedings{huang-etal-2026-uncertainty-aware,
  title={Uncertainty-Aware Planning for Disambiguating User Intent in Interactive LLM Agents: Application to Baidu Maps},
  author={Huang, Deqiang and Lu, Xinjiang and Zhou, Jingbo and Lu, Nijia and Li, Fuxin and Hong, Bo and Zhang, Chuanming and Xu, Tong and Chen, Enhong},
  booktitle={Proceedings of the 32nd ACM SIGKDD Conference on Knowledge Discovery and Data Mining},
  year={2026}
}

\clearpage
\appendix
\onecolumn

\section{Prompt Templates}
\label{app:prompts}

For reproducibility, we include the core prompt templates used in our closed-book clarify-then-search protocol.
Prompts A--C correspond to the Clarifier, User Answerer, and Rewriter, which define the interaction protocol.
Prompts D--E summarize the evidence-grounded nugget generation and coverage judging procedures.
The full executable prompts, JSON schemas, and scripts are released on the project page.

\vspace{0.5em}

\noindent
\begin{minipage}[t]{0.32\textwidth}
\footnotesize
\begin{tcolorbox}[
  title={Prompt A: Clarifier},
  colback=white,
  colframe=black!35,
  coltitle=black,
  colbacktitle=black!12,
  boxrule=0.4pt,
  arc=1mm,
  left=1.2mm,
  right=1.2mm,
  top=1mm,
  bottom=1mm,
  fonttitle=\bfseries\footnotesize,
  equal height group=evalprompts,   
  before upper={\setlength{\parskip}{2pt}}
]
\textbf{\# Role.}
You are a \emph{search clarification question generator}.

\textbf{\# Input.}
You are given a search-box query: \texttt{BLURRED\_QUERY}.
The query may be underspecified in time, location, target scope, criteria/definition, or other constraints.

\textbf{\# Task.}
(1) Identify key missing or ambiguous constraints only from \texttt{BLURRED\_QUERY}.
Do not introduce new entities, facts, or constraints not implied by the query.

(2) Select the top-\texttt{K} questions with the highest expected information gain.
Generate exactly \texttt{K} short clarification questions, ordered from highest to lowest priority.

(3) Each question must focus on one aspect only, such as time range, geographic region, entity disambiguation, scope, or evaluation criteria.
Questions must not be redundant or heavily overlapping.

(4) Do not rewrite or answer the query.
Output only the clarification questions.

\textbf{\# Output.}
Output a single JSON array with exactly \texttt{K} Chinese colloquial clarification question strings.
Do not output explanations, prefixes, suffixes, or code fences.
\end{tcolorbox}
\end{minipage}
\hfill
\begin{minipage}[t]{0.32\textwidth}
\footnotesize
\begin{tcolorbox}[
  title={Prompt B: User Answerer},
  colback=white,
  colframe=black!35,
  coltitle=black,
  colbacktitle=black!12,
  boxrule=0.4pt,
  arc=1mm,
  left=1.2mm,
  right=1.2mm,
  top=1mm,
  bottom=1mm,
  fonttitle=\bfseries\footnotesize,
  equal height group=evalprompts,   
  before upper={\setlength{\parskip}{2pt}}
]
\textbf{\# Role.}
You are acting as the \emph{real user}.
The user's true intent, i.e., the clear query, is provided to you privately.

\textbf{\# Inputs.}
The system provides:
\texttt{clear\_query} and one \texttt{clarification\_question}.

\textbf{\# Task.}
(1) Answer the clarification question only using information explicitly stated in \texttt{clear\_query}.
Do not fabricate or add any new facts or constraints.

(2) Answer only what the clarification question asks.
Do not provide additional information beyond the asked aspect.

(3) If the requested information is not explicitly specified in \texttt{clear\_query}, output exactly:
\texttt{unknown}.

(4) Keep the output extremely brief: a single short sentence or phrase.
No explanations, reasoning, or bullet points.

\textbf{\# Output.}
Output only the user answer text.
If the answer is not specified in \texttt{clear\_query}, output exactly \texttt{unknown}.
\end{tcolorbox}
\end{minipage}
\hfill
\begin{minipage}[t]{0.32\textwidth}
\footnotesize
\begin{tcolorbox}[
  title={Prompt C: Rewriter},
  colback=white,
  colframe=black!35,
  coltitle=black,
  colbacktitle=black!12,
  boxrule=0.4pt,
  arc=1mm,
  left=1.2mm,
  right=1.2mm,
  top=1mm,
  bottom=1mm,
  fonttitle=\bfseries\footnotesize,
  equal height group=evalprompts,   
  before upper={\setlength{\parskip}{2pt}}
]
\textbf{\# Role.}
You are acting as the \emph{search user}.
You only remember the original blurred query and the clarification Q\&A pairs.
You do not have access to the true intent query.

\textbf{\# Inputs.}
The system provides:
\texttt{blurred\_query} and \texttt{qa\_pairs}, where each pair contains a \texttt{clarification\_question} and a \texttt{user\_answer}.

\textbf{\# Task.}
(1) Rewrite the blurred query into a clearer, more specific, retrieval-ready single query using only \texttt{blurred\_query} and \texttt{qa\_pairs}.

(2) Do not introduce new entities, facts, or constraints not present in the inputs.

(3) If a \texttt{user\_answer} is \texttt{unknown}, do not force that aspect to be specific.
Keep it underspecified as in the original query.

(4) Output only the final rewritten query as a single sentence.
No explanations, steps, or multiple candidates.

\textbf{\# Output.}
Output exactly one rewritten query sentence in plain text.
\end{tcolorbox}
\end{minipage}

\vspace{0.6em}
\noindent\textbf{Figure 3:} Core closed-book interaction prompts. Prompt A generates clarification questions, Prompt B answers them under the hidden intent without adding extra information, and Prompt C rewrites the query without access to the hidden intent.


\section{Evaluation Prompt Summary}
\label{app:eval-prompts}

\noindent
\begin{minipage}[t]{0.49\textwidth}
\footnotesize
\begin{tcolorbox}[
  title={Prompt D: Evidence-Grounded Golden Reference Generator},
  colback=white,
  colframe=black!35,
  coltitle=black,
  colbacktitle=black!12,
  boxrule=0.4pt,
  arc=1mm,
  left=1.2mm,
  right=1.2mm,
  top=1mm,
  bottom=1mm,
  fonttitle=\bfseries\footnotesize,
  equal height group=evalprompts,   
  before upper={\setlength{\parskip}{2pt}}
]
\textbf{\# Role.}
You are a \emph{golden reference} generator for retrieval-augmented evaluation.

\textbf{\# Inputs.}
The input contains:
\texttt{QUERY}, a model \texttt{ANSWER}, and retrieved \texttt{SOURCES}, where each source contains identifiers and evidence snippets.

\textbf{\# Goal.}
Using only \texttt{SOURCES}, produce a traceable and scorable \texttt{GOLD} JSON.

\textbf{\# Mandatory Rules.}
(1) Use only \texttt{SOURCES} as factual grounding.
Do not introduce external knowledge or common sense.

(2) Decompose \texttt{QUERY} into sub-questions.
Only evidence-supported sub-questions may be included in \texttt{must\_cover}; unsupported requirements must be placed in \texttt{missing\_evidence}.

(3) Nuggets must be atomic and judgeable factual statements.
Each nugget must include one or more \texttt{source\_ids}.

(4) Compare \texttt{ANSWER} against \texttt{SOURCES} and list unsupported factual claims in \texttt{unsupported\_claims}.

(5) Output strict JSON only.

\textbf{\# Weighting Rule.}
\texttt{3}: core essential;
\texttt{2}: important information;
\texttt{1}: supplemental information.

\textbf{\# Main GOLD fields.}
\texttt{query\_intent}, \texttt{nuggets}, \texttt{reference\_urls}, \texttt{unsupported\_claims}, and \texttt{notes}.
Each nugget stores \texttt{id}, \texttt{text}, \texttt{weight}, and \texttt{source\_ids}.
\end{tcolorbox}
\end{minipage}
\hfill
\begin{minipage}[t]{0.49\textwidth}
\footnotesize
\begin{tcolorbox}[
  title={Prompt E: Nugget Coverage Judge},
  colback=white,
  colframe=black!35,
  coltitle=black,
  colbacktitle=black!12,
  boxrule=0.4pt,
  arc=1mm,
  left=1.2mm,
  right=1.2mm,
  top=1mm,
  bottom=1mm,
  fonttitle=\bfseries\footnotesize,
  equal height group=evalprompts,   
  before upper={\setlength{\parskip}{2pt}}
]
\textbf{\# Role.}
You are a rigorous \emph{coverage judge} for nugget-based evaluation.

\textbf{\# Inputs.}
The input contains:
\texttt{QUERY}, \texttt{GOLD\_NUGGETS}, and \texttt{CANDIDATE\_ANSWER}.

\textbf{\# Task.}
For each golden nugget, decide whether \texttt{CANDIDATE\_ANSWER} covers the nugget's meaning.
Judging must be strict and based only on the candidate answer.

\textbf{\# Labels.}
\texttt{full}: semantically consistent and includes all critical details.

\texttt{partial}: mentions the core idea but misses a critical qualifier, object, mechanism, or condition.

\texttt{none}: not mentioned, contradicted, or only irrelevant content is present.

\textbf{\# Rules.}
Do not award credit using external knowledge or what should be true.
Judge only what is explicitly present in \texttt{CANDIDATE\_ANSWER}.
If the answer contradicts a nugget, label it \texttt{none}.

\textbf{\# Output.}
Output strict JSON parseable by \texttt{json.loads}:
\texttt{\{"results":[\{"id":"N1","coverage":"full"\}, ...]\}}.
Do not output any additional text.
\end{tcolorbox}
\end{minipage}

\noindent\textbf{Figure 4:} Evaluation prompts. Prompt D constructs evidence-grounded static golden nuggets, while Prompt E judges nugget coverage using \texttt{full}/\texttt{partial}/\texttt{none} labels. Full executable prompts and JSON schemas are released with the artifact.

\clearpage
\section{Additional Diagnostics}
\label{app:additional-diagnostics}

This appendix provides additional diagnostics omitted from the main text due to space limits. Table~10 gives the full turn-level region-only statistics behind the compact summary in Table~8. Table~11 summarizes supporting-backbone stability for the UA and Rewriter components.

\begin{table}[h]
\centering
\scriptsize
\setlength{\tabcolsep}{4pt}
\renewcommand{\arraystretch}{0.92}
\caption{Turn-level UA \texttt{unknown} rate conditioned on region-only questions on the benchmark set ($n=518$).}
\label{tab:app_regiononly_turn}
\begin{tabular}{llcc}
\toprule
$k$ & (Model, turn) & \#region-only & region-only unknown \\
\midrule
\multirow{7}{*}{$1$}
& (Qwen, 1)     &  64 & 0.766 \\
& (ERNIE, 1)    & 136 & 0.787 \\
& (DeepSeek, 1) & 223 & 0.807 \\
& (Kimi, 1)     &  48 & 0.750 \\
& (GPT, 1)      & 121 & 0.785 \\
& (Claude, 1)   & 148 & 0.824 \\
& (Gemini, 1)   & 236 & 0.869 \\
\midrule
\multirow{14}{*}{$2$}
& (Qwen, 1)     &  71 & 0.831 \\
& (Qwen, 2)     &  24 & 0.917 \\
& (ERNIE, 1)    &  54 & 0.852 \\
& (ERNIE, 2)    &  38 & 0.868 \\
& (DeepSeek, 1) & 239 & 0.841 \\
& (DeepSeek, 2) &  61 & 0.885 \\
& (Kimi, 1)     &  90 & 0.822 \\
& (Kimi, 2)     &  23 & 0.870 \\
& (GPT, 1)      & 108 & 0.787 \\
& (GPT, 2)      &  24 & 0.833 \\
& (Claude, 1)   & 198 & 0.838 \\
& (Claude, 2)   &  22 & 0.818 \\
& (Gemini, 1)   & 222 & 0.887 \\
& (Gemini, 2)   &  25 & 0.800 \\
\midrule
\multirow{21}{*}{$3$}
& (Qwen, 1)     &  52 & 0.769 \\
& (Qwen, 2)     &  35 & 1.000 \\
& (Qwen, 3)     &  12 & 0.917 \\
& (ERNIE, 1)    &  67 & 0.881 \\
& (ERNIE, 2)    &  51 & 0.882 \\
& (ERNIE, 3)    &  33 & 0.879 \\
& (DeepSeek, 1) & 231 & 0.879 \\
& (DeepSeek, 2) &  74 & 0.824 \\
& (DeepSeek, 3) &   6 & 1.000 \\
& (Kimi, 1)     &  94 & 0.830 \\
& (Kimi, 2)     &  33 & 0.788 \\
& (Kimi, 3)     &  13 & 0.692 \\
& (GPT, 1)      & 113 & 0.779 \\
& (GPT, 2)      &  27 & 0.778 \\
& (GPT, 3)      &   9 & 0.889 \\
& (Claude, 1)   & 203 & 0.852 \\
& (Claude, 2)   &  25 & 0.920 \\
& (Claude, 3)   &  15 & 1.000 \\
& (Gemini, 1)   & 212 & 0.877 \\
& (Gemini, 2)   &  19 & 1.000 \\
& (Gemini, 3)   &   7 & 1.000 \\
\bottomrule
\end{tabular}
\end{table}

\vspace{0.8em}

\begin{table}[h]
\centering
\scriptsize
\setlength{\tabcolsep}{4pt}
\renewcommand{\arraystretch}{0.92}
\caption{Supporting-backbone stability summary. UA agreement measures whether two supporting backbones produce the same known/unknown status; rewrite similarity is normalized text similarity between rewritten queries; unknown preservation measures whether rewrites avoid concretizing unknown answers.}
\label{tab:app_pairwise_stability}
\begin{tabular}{lccc}
\toprule
Pair & UA agree & Rewrite sim. & Unknown preserv. \\
\midrule
GPT-ERNIE vs. GPT-GPT     & 0.865 & 0.771 & 0.988 \\
GPT-ERNIE vs. GPT-Claude  & 0.874 & 0.709 & 0.987 \\
GPT-GPT vs. GPT-Claude    & 0.881 & 0.726 & 0.990 \\
ERNIE-ERNIE vs. ERNIE-GPT    & 0.885 & 0.798 & 0.989 \\
ERNIE-ERNIE vs. ERNIE-Claude & 0.880 & 0.745 & 0.990 \\
ERNIE-GPT vs. ERNIE-Claude   & 0.891 & 0.756 & 0.994 \\
\bottomrule
\end{tabular}
\end{table}
\end{document}